\documentclass[]{acta}
\usepackage{supertabular,lscape,epsfig}
\usepackage{amssymb}
\usepackage{amsmath}
\usepackage{multirow}

\SetPages{0}{0}

\SetVol{71}{2021}

\begin{document}

\begin{Titlepage}
\Title{A Detection Threshold in the Amplitude Spectra Calculated from {\it TESS} Time-Series Data}
\Author{A.S. B~a~r~a~n$^{1,2,3}$, C. K~o~e~n$^{4}$}
{
$^{1}$ARDASTELLA Research Group, Institute of Physics, Pedagogical University of Cracow, ul. Podchor\c{a}\.zych 2, 30-084 Krak\'ow, Poland\\
$^{2}$Embry-Riddle Aeronautical University, Department of Physical Science, Daytona Beach, FL\,32114, USA\\
$^{3}$Department of Physics, Astronomy, and Materials Science, Missouri State University, Springfield, MO\,65897, USA\\
$^{4}$Department of Statistics, University of the Western Cape, Bellville 7535, Cape Town, South Africa\\
e-mail:andrzej.baran@up.krakow.pl
}

\Received{Month Day, Year}
\end{Titlepage}

\Abstract
{We present results of time-series data simulation. We aimed at estimating the threshold used for detecting signals in amplitude spectra, calculated from simulating {\it TESS} photometry of up to one year duration. We selected the threshold at a false alarm probability FAP\,=\,0.1\% and derived S/N ratios between 4.6 and 5.7 depending on the data cadence and coverage. We also provide a formula to estimate the threshold for any FAP adopted and a given number of data points. Our result confirms that, to avoid spurious detection, space-based photometry may require substantially higher S/N than that typically being employed for ground-based data.}
{stars: variable stars; Fourier technique}

\section{Introduction}
Time-series data of variable stars are commonly analyzed by means of the Fourier technique. As a first step in any analysis an amplitude or power spectrum (amplitude or power vs frequency $\nu$) is calculated. The power spectrum, or periodogram, is defined as
\begin{equation}
I(\omega) = \frac{1}{N_p} \left \{ \left [ \sum_t (y_t-\overline{y})
\cos \omega \,t \right ]^2+\left [ \sum_t (y_t-\overline{y})
\sin \omega \,t\right ]^2 \right \} \;\;\;\;\;  0<\omega<\pi
\end{equation}
where $N_p$ is the series length, $y_t$ is the $t$-th measurement, $\overline{y}$ is the 
mean of the measurements, and $\omega=2\pi \nu$ is the angular frequency (Press et al. 2007).
The amplitude spectrum is
\begin{equation}
S(\omega)=2\sqrt{\frac{I(\omega)}{N_p}}
\end{equation}
In this paper the focus is on $S(\omega)$; it is not difficult to
obtain corresponding results for $I(\omega)$ by transformation (see e.g.
Koen 2015).

The spectrum maximum, i.e. the largest peak, may be indicative of the presence of a periodicity in the data. The long standing concern is about the significance of the largest peak, since the observations always contain noise that can induce spurious peaks. If an amplitude of the peak is much larger than the noise level it looks convincing to be accepted as a real signal. As we approach low amplitude peaks we have more doubts. That is why a certain condition, commonly called a detection threshold, comes into play, and should work as a discriminant between the real signal and noise. This condition is a statistic, which depends on the reliability with which one wants to avoid a spurious detection. The threshold is often defined as a multiple of the signal to noise (S/N) ratio and, in case of ground-based data, commonly accepted as four (e.g. Breger et al.\,1993 -- but see Koen\,2010). (Note that throughout the paper we will be referring to the ratio of the highest spectrum peak to the average level of the spectrum as the ``S/N" ratio, regardless of whether there is in fact any signal in the time series. In what follows the ``average" value of the spectrum will be taken to be the median. Percentiles given below can easily be transformed to the equivalent for mean-standardised spectra by using Eq.\,(A6). The threshold of S/N\,=\,4 cannot be uncritically applied to continuous space-based data of long coverage. The probability of a spurious detection is higher and we can anticipate that the threshold should be larger. In fact, Baran, Koen \& Pokrzywka\,(2015) simulated the threshold in the case of {\it Kepler} K2 short cadence data resulting in higher S/N than for ground-based data. The authors also provided discussion and literature references regarding detection of a signal.

In this work we aim at estimating the detection threshold applicable to time-series data sampled at the ultra-short ($\Delta\,t$=\,20\,sec), short ($\Delta\,t$=\,120\,sec) and long ($\Delta\,t$=\,1800\,sec) cadences provided by the Transiting Exoplanet Survey Satellite ({\it TESS}, Ricker et al.\,2015). We use a false alarm probability (FAP), which gives the probability that we will be wrong claiming signal detection. Therefore, our work is consistent with previous efforts characterised by specifying the null hypothesis H0, {\it there is no signal in the data}. The detection threshold is not a deterministic quantity. It depends on how certain one wants to be to avoid detection of false positives. Therefore, first we need to specify the FAP and then derive a corresponding amplitude or S/N ratio. In our work we adopt the FAP\,=\,0.1\%

The goal of this work is to provide S/N ratios corresponding to the given FAP, and number of ``sectors" (i.e. individual observing runs) by {\it TESS}. We expect that the results of our simulations will be widely applicable and will help others to save computation time and, more importantly, to avoid false detections in amplitude spectra. 

\section{Simulations}
We used {\sc python} to simulate the ultra-short, short and long cadence {\it TESS} time series data sets of Gaussian noise. There are two distinct issues raised by the assumption that the noise consists of uncorrelated Gaussians. One is the possibility of trends in the data. Those can be dealt with by detrending or by avoiding, if possible, frequencies where the power is elevated due to trends. The second issue is whether noise is Gaussian, or whether it has some other distribution. For datasets of the length considered here, this is immaterial -- the distribution of the spectrum is largely impervious to the noise distribution (Koen 2021). {\it TESS} data are delivered in sectors, where one sector contains data collected over two {\it TESS} orbits. Depending on Galactic latitude a star can be observed during a single sector, several sectors, or as many as 13 sectors. The latter targets are located in the ``continuous viewing zone". The average level in an amplitude spectrum depends on both the root-mean-square (RMS) of the Gaussian noise and the number $N_p$ of data points in the time-series data. The dependence on the former is easily dealt with by standardising the spectrum, i.e. dividing by e.g. its mean value -- it is clear from Eq.\,(1) and (2) that in the case of pure noise $y_t$, the RMS is merely a scale factor, which is eliminated by using a suitable quotient. In the simulation results reported in Section\,3 we choose the spectrum median for standardisation, since this is generally more resistant to outlying values (as might be induced by the presence of signals in the data).

In order to model how the detection threshold depends on $N_p$ we run our simulations for a number of sectors between 1 and 13, keeping in mind that short gaps, caused by {\it e.g.} a downlink or technical issues, have negligible influence on the results of our analysis. Each data set, for a specific number of sectors, contains a multiple of a number of points in one sector, which is $N_p=116\,640, 19\,440, 1296$, for the ultra-short, short and long cadence, respectively. We used a fast Fourier algorithm to calculate amplitude spectra of simulated time-series data. The frequency ranges up to the Nyquist frequency of $\nu_{Nyq}=0.5/{\Delta t}=2160, 360, 24$ c/d, respectively for the ultra-short, short and long cadences. A regular frequency grid is used in Eq.\,(2). The frequency resolution of the spectrum is of the order of $1/T$, where $T$ is the total time (including gaps) spanned by the observations. This ranges from 27\,days for one sector data, to 351\,days for 13 sector data. The corresponding resolutions range from 0.037 to 0.0028 c/d. We oversample the spectrum by a factor of 9 in order to ensure that peaks are fully resolved; frequency steps are in the range 0.0041\,--\,0.00032\,c/d.

\section{Results}
The number of simulated datasets with a detection of a peak with a specific amplitude decreases as the amplitude increases. This is expected since the chance to find larger amplitude noise-induced peaks is smaller. We estimated the S/N values for an arbitrarily chosen RMS and for data lengths of 1 to 13 sectors. In Fig.\,1 we plot the S/N as a function of the number of sectors for the three cadences. The curves can be described by logarithmic functions:
\begin{eqnarray}
S/N &=&5.2902(48)\,+\,0.1351(26) \cdot \ln N_s~~~{\rm for}~~~\Delta t=20\,sec\nonumber\\
S/N &=&5.0355(38)\,+\,0.1417(20) \cdot \ln N_s~~~{\rm for}~~~\Delta t=120\,sec\\
S/N &=&4.6200(29)\,+\,0.1559(15) \cdot \ln N_s~~~{\rm for}~~~\Delta t=1800\,sec\nonumber
\end{eqnarray} 
where $N_s$ is the number of sectors. In Table\,1 we list the S/N ratio we derived for a given data coverage. The S/N ratio clearly shows that the value of four, commonly accepted for ground based data, is too optimistic and when applied to space based continuous data, many large noise induced peaks are detected in amplitude spectra. In the case of the most common {\it TESS} short cadence data, a S/N ratio of 5 and higher is required. For the ultra-short cadence data, the threshold rises to $\sim 5.3$. Obviously, if one is interested in FAP higher than 0.1\%, the S/N threshold will be lowered. 

\MakeTable{|cc|ccc|}{12.5cm}{S/N at FAP\,=\,0.1\% for the ultra-short, short and long cadence times-series data.}
{\hline
&& \multicolumn{3}{c|}{S/N} \\
&&20\,sec & 120\,sec & 1800\,sec\\
\hline
\multirow{13}{*}{\rotatebox[origin=c]{90}{Number of sectors}}
& 1 & 5.292 & 5.037 & 4.615 \\
& 2 & 5.392 & 5.124 & 4.731 \\
& 3 & 5.433 & 5.194 & 4.790 \\
& 4 & 5.468 & 5.238 & 4.835 \\
& 5 & 5.497 & 5.266 & 4.877 \\
& 6 & 5.538 & 5.287 & 4.890 \\
& 7 & 5.555 & 5.321 & 4.924 \\
& 8 & 5.575 & 5.330 & 4.947 \\
& 9 & 5.585 & 5.342 & 4.962 \\
& 10 & 5.604 & 5.362 & 4.981 \\
& 11 & 5.622 & 5.369 & 4.998 \\
& 12 & 5.628 & 5.391 & 5.002 \\
& 13 & 5.629 & 5.397 & 5.013 \\
\hline
}

As could be expected, for a given data coverage, the chance of a random noise-induced peak is higher, if the number of data points increases, and therefore the S/N ratio for the ultra-short cadence data is higher, while it is lower for the long cadence data, {\it e.g.} the S/N ratio for 15 sectors of  long cadence data is similar to the ratio for one sector of short cadence data, while the S/N ratio for 6 sectors of short cadence data is about the same as that for one sector of ultra-short cadence data.

We expect that sector-long gaps in the case of multi-sector data may change the S/N threshold at a given FAP. If we compare two sector data with three sector data but having a sector gap in the middle, the number of observations remains the same but the frequency resolution is different. In the latter case the number of calculated frequencies will be higher, which will increase the odds for spurious detection and hence the S/N should be higher. Our simulations confirm our expectation. The difference depends on the exact configuration, i.e. the positions and lengths of any gaps in the time-series data. It should also be noted that in the case of long gaps aliasing may arise, i.e. the spectral peaks at widely separated frequencies may no longer be independent. 

\begin{figure}[htb]
\includegraphics{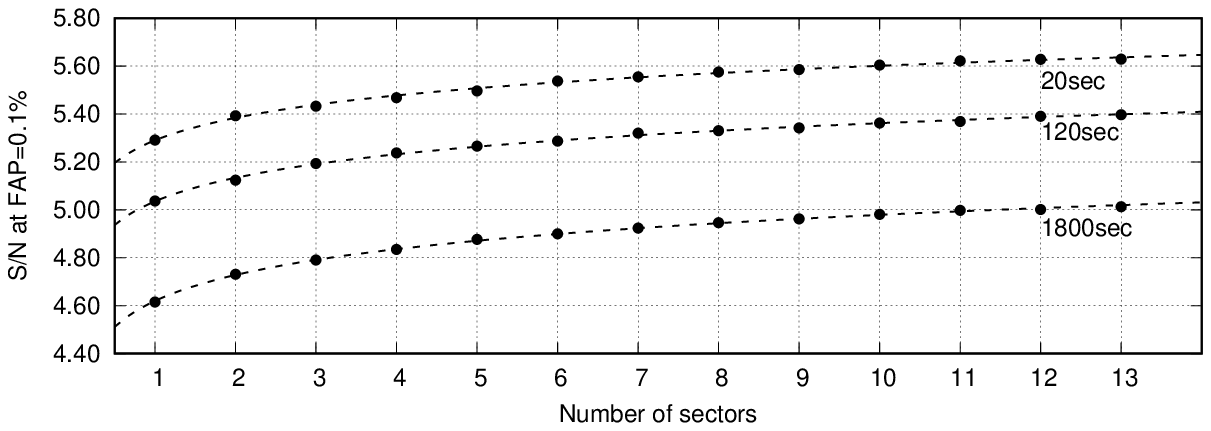}
\FigCap{S/N ratio as a function of time series length for the three {\it TESS} observing cadences.}
\end{figure}

\section{Some further considerations}
In Appendix we derive expressions for the cumulative distribution functions (CDFs) of maxima of the amplitude spectrum. These can be used in turn to find equations for percentiles (S/N values) for given FAPs. Two standardisations are considered -- division of the spectrum maximum by the spectrum mean, and division of the spectrum maximum by the spectrum median. These give rise to statistics denoted $W$ and $U$ in Appendix [see Eq.\,(A2) and (A5)].

Percentiles are found by setting the CDF equal to the required probability and solving for the statistic:
\begin{equation}
1-F_U(u_p)=p
\end{equation}
i.e. the probability that $U>u_p$ is $p$. For $F_U$ taken from Eq.\,(A8) and (A7),
\begin{eqnarray}
u_p&=&\frac{2c}{\sqrt{\pi}}\left \{ \mu\ln N_p -\sigma \ln [-\ln~(1-p) 
\right \}^{1/2}\nonumber\\
 &=& 1.201 \left \{ 1.05 \ln N_p -1.04 \ln [-\ln~(1-p)] \right \}^{1/2}
\end{eqnarray}
[see Eq.\,(A1)]. For $p=0.001$,
\begin{equation}
u_{0.001}=1.201 \sqrt{1.05\ln N_p+7.184} 
\end{equation}
Fig.\,2 compares the simulation results listed in Table\,1 (obtained for FAP\,=\,0.1\%) with the prediction of equation (6): the agreement is evidently quite good. The figure also shows the predictions for three other FAPs, i.e. 0.01, 1 and 5\%.

The results presented up to this point presuppose that the amplitude spectrum over the entire frequency interval $(0,\nu_N]$ is tested for significant features. In practice, one may only be interested in narrow frequency intervals corresponding to the periods of specific types of variable stars. For example, if searching for High Amplitude Delta Scuti stars, one may want to restrict the frequency range to say 5-21\,days$^{-1}$ (e.g. McNamara\,2000). This represents fractions $f=0.0074, 0.044, 0.67$ of the range $(0,\nu_{Nyq}]$ for the ultra-short, short and long cadences, respectively. The implication is that in this case the data length $N_p$ needs to be replaced by $N_{\rm eff}=N_p f$ when determining the appropriate threshold for a given FAP. For a single sector and the frequency range quoted, $N_{\rm eff}$ is 864, which changes the $p=0.001$ threshold to 4.54 for all three cadences. This is significantly smaller than 5.04 and 5.29 for the full frequency ranges for the short and ultra-short cadences respectively.

\begin{figure}[htb]
\includegraphics{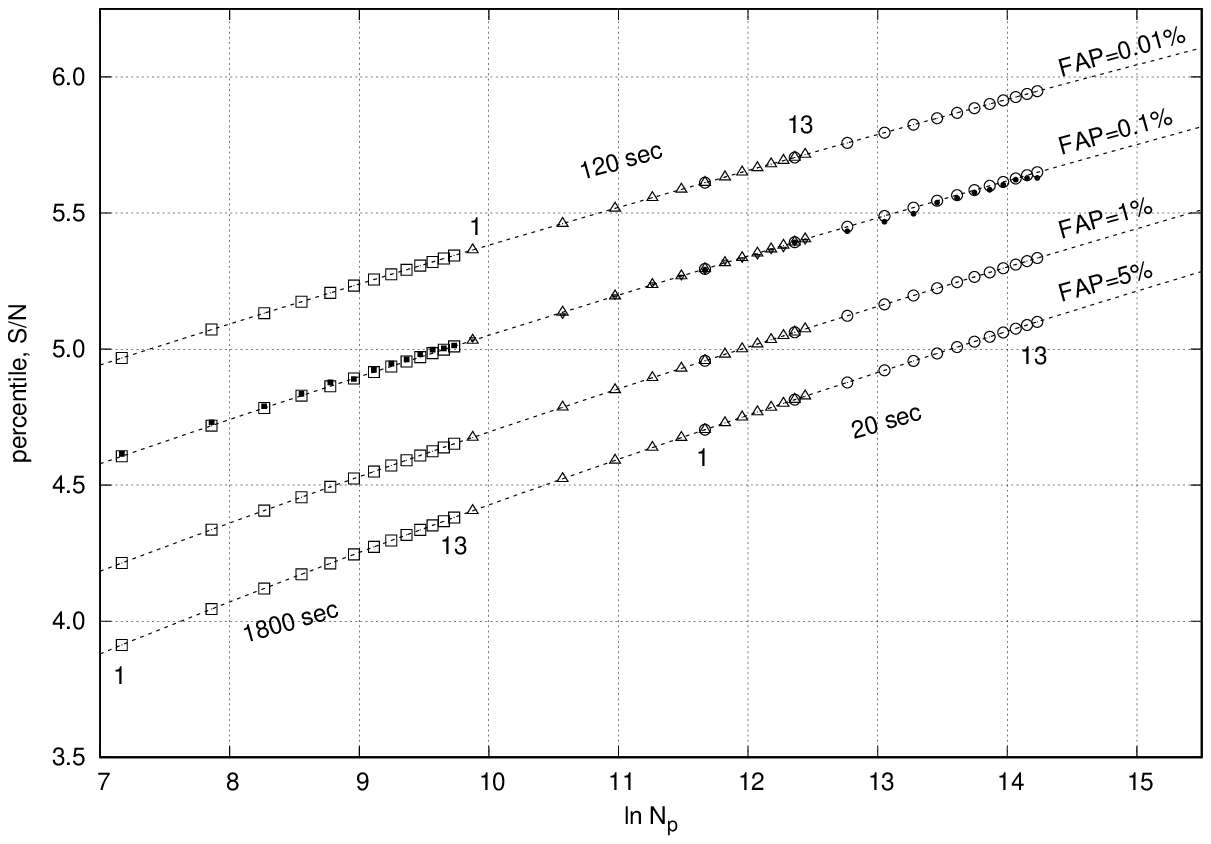}
\FigCap{A plot showing percentiles for four given FAPs, calculated from equation (5), in function of number of data points for the three {\it TESS} cadence data. Open circles show the ultra-short cadence data, open triangles show the short cadence data and open squares show the long cadence data. The labels 1 and 13 indicate the number of sectors. The filled symbols show the values from Table\,1. As explained in the text, the points for 1 and 2 sector ultra-short cadence data overlap with 6 and 12 sector short cadence data. The dashed lines show the general trend of equation (5) for four given FAPs.}
\end{figure}

\section{Summary}
We presented the result of our simulations designed to derive the S/N ratio at a FAP\,=\,0.1\%. The threshold we focused on implies that spectrum peaks are only this large by chance 0.1\% of the time. We ran 10\,000 simulations for a given RMS, data coverage and ultra-short, short and long cadences, used to collect data by the {\it TESS} satellite. Since the average noise level in an amplitude spectrum is proportional to the RMS and inversely proportional to the square root of the number of data points, we could anticipate the result for a selection of input parameters. According to our results, the S/N ratio at FAP\,=\,0.1\% ranges between 5.3 and 5.63, 5.0 and 5.4, 4.6 and 5.0, for the ultra-short, short and long cadence time-series data, respectively. If sectors-long gaps are involved, the threshold is higher, even by 0.25, as compared to the gap-free time-series data with the same number of points. This result confirms our expectation that more data points and longer data coverage, than we typically achieve from the ground, increases the likelihood of noise-induced peaks and therefore S/N\,=\,4, commonly applied to the ground-based data, is generally inadvisable in the case of space data.

The material in this paper is subject to a number of caveats:
\begin{itemize}
\item[(1)]
If many datasets are examined at a specific FAP, the overall $p$-value is increased. This can be guarded against by suitable false discovery procedures (e.g. Benjamini \& Hochberg\,1995).
\item[(2)]
The presence of many large amplitude signals in a time-series data would inflate the average noise level and alter the distribution of the peak values.
\item[(3)]
The test procedure is designed to be applied once to a given dataset. Prewhitening a time-series data and testing the residuals could lead to suspect results, for at least two reasons: first, the multiple testing effect referred to in (1), and second, the effects of prewhitening on the spectrum of the residuals (see e.g. Koen\,2010 simulations suggest that, on average, prewhitening removes an excessive amount of power from the spectrum. This could be mitigated by ``partial" prewhitening, as in the CLEAN algorithm (see e.g. Stoica \& Moses 2005, Section 6.5.7).
\item[(4)]
Observations acquired by space-based telescopes often suffer from instrumental effects which manifest as excess power at low frequencies in spectra. It may be necessary to avoid the lowest frequencies and to only work with spectra over $[\nu_0,\nu_{Nyq}]$, where $\nu_0$ is some suitable cutoff frequency. This will have implications for the effective number of observations $N_p$ (see the discussion at the end of section\,4).
\item[(5)]
The effects of finite exposure times on the spectrum have not been taken into account.  
\item[(6)]
The material in this paper addresses testing for significance the largest peak in a spectrum calculated over a pre-specified frequency interval. In the case of a single pre-specified frequency, the fact that individual periodogram ordinates are exponentially distributed (e.g. Frescura et al. 2008 can be used to construct significance tests).
\end{itemize} 

These caveats apply to {\it all} analysis methods, not only those presented in this paper, but are rarely acknowledged. In practical applications (1) and (3) are probably the most serious, as variable star time-series data are often subject to several cycles of prewhitening. The remedies suggested above should be useful in this regard, but further research is needed.

\Acknow{Financial support from the National Science Centre under projects No.\,UMO-2017/26/E/ST9/00703 and UMO-2017/25/B/ST9/02218 is acknowledged.}

\pagebreak

\appendix
\section*{\large \bf Appendix: Distributions of standardised amplitude spectra.}

 \setcounter{section}{0}
 \def\thesection{A\arabic{section}}
 \setcounter{equation}{0}
 \def\theequation{A\arabic{equation}}

Koen\,(2015) found that for the periodogram defined in (1) the standardised maximum $$V=\max_\omega I(\omega)/\overline{I(\omega)}$$ has, to good approximation, the Gumbel distribution with probability density function (PDF) $$f_V(v)=(1/\sigma)\exp \left \{ -\frac{(v-\mu)}{\sigma}-\exp \left [-\frac{(v-\mu)}{\sigma} \right ] \right \}$$ where 
\begin{equation}
\sigma \approx 1.04 \;\;\;\;\;\;\ \mu \approx 1.05 \ln N_p
\end{equation}
Then it can be shown that the standardised amplitude spectrum maximum
\begin{equation}
W=\max_\omega S(\omega)/\overline{S(\omega)}
\end{equation}
has the PDF of the following form
\begin{eqnarray}
f_W(w)&=&\frac{\pi}{2\sigma} \exp \left \{ -g(w)-\exp \left [-g(w) \right ] \right \} \nonumber\\
g(w) &=& (\pi w^2-4\mu \ln N_p)/(4\sigma) 
\end{eqnarray}
with $\mu$ and $\sigma$ as in Eq.\,(A1). The cumulative distribution function (CDF) corresponding to the PDF given in Eq.\,(A3) is
\begin{equation}
F_W(w)=\exp \left \{ -\exp [-g(w)] \right \} \; .
\end{equation}

An alternative standardisation of the amplitude spectrum is to divide by the spectrum median $\widetilde{S(\omega)}$, i.e. the statistic of interest is
\begin{equation}
U=\max_\omega S(\omega)/\widetilde{S(\omega)}\; .
\end{equation}
Koen\,(2015) pointed out that $S(\omega)$ as defined in Eq.\,(2) follows the Rayleigh distribution;  the ratio of mean to median for this distribution is
\begin{equation}
c=\overline{S(\omega)}/ \widetilde{S(\omega)}=\sqrt{\frac{\pi}{4\ln(2)}}=1.0645 \;.
\end{equation}
It follows from Eq.\,(A2) that
\begin{eqnarray}
f_U(u)&=&\frac{1}{c}f_W(u/c)\nonumber\\
&=&\frac{\pi u}{2c^2\sigma} \exp \left \{ -h(u)-\exp \left [ 
-h(u) \right ] \right \} \nonumber\\
h(u) &=& (\pi u^2-4\mu c^2\ln N_p)/(4\sigma c^2)  
\end{eqnarray}
with corresponding CDF
\begin{equation}
F_U(u)=\exp \left \{ -\exp [-h(u)] \right \} \; .
\end{equation}

%\multicolumn{7}{p{9cm}}{All tables should be set in {\TeX}/{\LaTeX}. Each table should be numbered sequentially with Arabic numerals and be supplemented by a short title describing its content.}

\end{document}